\documentclass[%
 reprint,
 amsmath,amssymb,
 aps,
floatfix,
]{revtex4-2}

\usepackage{graphicx}
\usepackage{dcolumn}
\usepackage{bm}
\usepackage{multirow}
\usepackage{siunitx}
\usepackage[caption=false]{subfig}

\begin{document}

\preprint{APS/123-QED}

\title{Angular Resolution of Electrons in Gaseous Targets}

\author{Majd Ghrear}
\email{majd@hawaii.edu}
\author{Sven E. Vahsen}%
\email{sevahsen@hawaii.edu}
\affiliation{%
Department of Physics and Astronomy, University of Hawaii,\\2505 Correa Road, Honolulu, HI, 96822, USA
}%

\date{\today}

\begin{abstract}
Low-energy electron recoils are of interest in several planned and proposed future nuclear and particle physics experiments. The topology and directions of such recoils provide important particle identification and kinematical constraints, and are experimentally accessible in gaseous targets. Electron recoils have complex trajectories, and the angular resolution that can be achieved has not been well understood. We have developed a method for estimating and optimizing this angular resolution, considering contributions from both multiple scattering and detection. First, we clarify that the formula commonly used for multiple scattering through small angles is actually a fit to Moliere theory for heavy particles. We revise this formula so that it is applicable to electrons in gas. Next, we combine this with an effective point resolution contribution, which accounts for diffusion and detector effects, to obtain an approximation for the angular resolution. We identify the optimal fit length and the corresponding optimal angular resolution. The result is a simple formula to estimate the best achievable angular resolution for electrons in gaseous detectors, given the electron energy and basic gas and detector properties. Our model's predictions show good agreement with simulations. This approach can assist in the design of future experiments and the development of analysis techniques. Given the widespread use of gaseous detectors, this work is relevant to many scientific communities.
\end{abstract}

\maketitle


\section{\label{sec:level1} Introduction}

When a free electron traverses a gaseous medium, it undergoes multiple scattering, altering its trajectory and generating a track of secondary, ionized electrons. This fundamental process plays a crucial role in various fields of physics, influencing beam halo formation at particle accelerators~\cite{PhysRevAccelBeams.21.051001}, dose rate calculations in medical physics~\cite{https://doi.org/10.1002/mp.16697}, and signal degradation in near-ambient pressure X-ray photoemission spectroscopy~\cite{https://doi.org/10.1002/sia.6947}. It is also essential for detecting electron tracks in gaseous targets used in rare-event searches.

An electron recoil occurs when an incident particle---such as a gamma ray, neutrino, or dark matter particle---transfers energy to an atomic electron. This electron then propagates through the target medium, producing a cloud of secondary ionization that may be fully contained in the target. In the low-energy regime ($1-100$\,keV) electron recoils constitute a dominant background in weakly interacting massive particle (WIMP) dark matter searches.  Developing strategies to reject electron recoils is therefore a major challenge for future experiments~\cite{PhysRevLett.132.111801}. Recently, however, electron recoils have been recognized as a potential signal for new dark matter interactions, including those predicted via the Migdal effect~\cite{PhysRevD.108.072006}.

The topology of electron recoils contain valuable information, yet remains difficult to measure. The recoil length and charge density can help distinguish electron recoils from nuclear recoils~\cite{Ghrear:2020pzk,battat2014radon,billard2012low,lopez2012background,dinesh,MIMAC2,CYGNO_erej}. Furthermore, the directions of nuclear recoils can provide evidence for the galactic origin of dark matter and help discriminate between dark matter and solar neutrinos~\cite{Vahsen:2021gnb, Vahsen:2020pzb,Battat:2016pap,Mayet:2016zxu,OHare:2021utq}.

Recently, interest in directional electron recoil detection has increased. The Imaging X-ray Polarimetry Explorer (IXPE)~\cite{weisskopf2016imaging} uses electron directionality to measure astrophysical X-ray polarization. The directional detection of electron recoils in gaseous detectors also has potential for solar neutrino measurements~\cite{Lisotti:2024fco,Torelli:2024mof}. Additionally, the MIGDAL experiment~\cite{MIGDAL:2022yip} aims to precisely measure the directions of overlapping nuclear and electron recoil tracks.

Detecting the directions of low-energy electron recoils requires gaseous detectors, where the recoil length exceeds the diffusion scale of the secondary ionization. Modern micro-pattern gaseous detectors (MPGDs) offer the necessary spatial resolution and sensitivity to reconstruct recoil topology down to the keV scale~\cite{Phan_2020}. In gaseous time projection chambers with MPGD-based readouts, the  ionization trail can be reconstructed in three dimensions~\cite{Ghrear:2024rku}.

Two primary factors limit the angular resolution of electron recoils in gaseous detectors: the multiple scattering of the primary electron and the detector resolution for secondary ionization measurements. Somewhat surprisingly, we find that the commonly used formula for multiple scattering through small angles fails to accurately describe electron recoils at the 100 keV scale. Given the growing interest in utilizing electron recoils for fundamental physics, an updated scattering formula is critical for optimizing future experiments. Additionally, it is necessary to understand how detector resolution and reconstruction techniques impact angular resolution. Since the complex topology of electrons recoils makes them challenging to analyze, the method of direction reconstruction also significantly affects the achievable resolution. Both conventional and machine learning approaches have been explored for this purpose~\cite{Torelli:2024mof,DiMarco_2022,Ghrear:2024rku}.

 To address these challenges, we develop a convenient and practical method for estimating the angular resolution of electrons in an arbitrary gas mixture. This solves the three problems outlined above. First, in Section~\ref{MS}, we develop a small-angle approximation for the multiple scattering of electrons. Second, in Section~\ref{PR}, we incorporate detector resolution effects. Finally, in the same section, we introduce a simple formula for predicting the best achievable angular resolution for electrons in gas. 

Our formalism depends on basic gas properties and the detector's effective point resolution, assuming an idealized detector capable of secondary-electron counting. Modern gaseous detectors are approaching this fundamental  limit~\cite{Ligtenberg:2020ofy,Ghrear:2024bxi,Sorensen:2012qc}, making our predictions relevant for current and future experiments.

Our method enables rapid optimization of detector designs by providing a fast and accurate way to estimate electron recoil angular resolution. Since future gaseous detectors will consider many gas mixtures, quick performance estimates can streamline the design process. While full simulations and experimental validation remain essential, our framework sets an upper bound on achievable performance and provides a valuable tool for guiding detector development.

\section{Multiple Scattering through Small Angles}
\label{MS}

\subsection{Background}

\begin{figure}[b]
\includegraphics[width=0.49\textwidth]{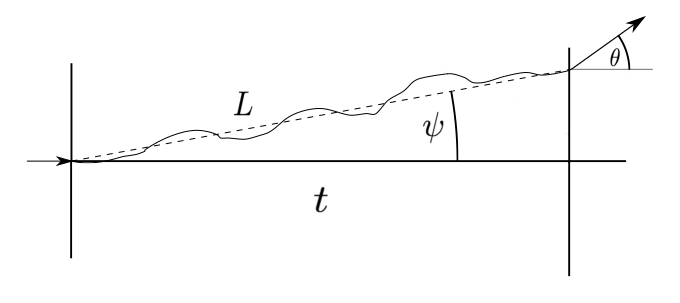}
\caption{\label{fig:MS_pic} An illustration of multiple scattering projected onto a plane containing the initial direction. $L$ is the point-to-point distance between the start and end of the track, illustrated by the dashed line, $t$ is the thickness of the material, $\theta$ is the angle to the initial direction upon exiting the material, and $\psi$ is the angle from the initial direction to the linear path within the material.}
\end{figure}

Rossi and Greisen were among the first to study multiple scattering through small deflections, proposing the simple approximation~\cite{RevModPhys.13.240}
\begin{equation}
 \sigma_\psi^{ \rm plane} = \frac{1}{\sqrt{3}} \frac{ \SI{14.8}{MeV} } { \beta c p} \sqrt{\frac{x}{X_o}}.
 \label{Rossi_MS}
\end{equation}
Above, $\sigma_\psi^{ \rm plane}$ is the RMS of the projected angle $\psi$ (illustrated in Fig.~\ref{fig:MS_pic}), $\beta$ and $p$ are the speed and momentum of the particle, and $x/X_o$ is the thickness of the scattering medium in radiation lengths. Rossi and Greisen arrive at their formula via statistical methods, where the elementary scattering probability is obtained by considering elastic scattering off of a fixed point charge nucleus. Slight modifications are included to account for the finite size of the nucleus and the screening of the nuclear field by outer electrons. Within their approximations, the spin-dependence is negligible, resulting in a general formula that is applicable to any particle with unit charge. As stated, Eq.~\ref{Rossi_MS} assumes negligible energy loss; however, Ref.~\cite{RevModPhys.13.240} discusses how energy loss can be incorporated.

Later, Moliere theory~\cite{Moliere} was developed which could be used to calculate detailed, non-Gaussian distributions for multiple Coulomb scattering through small angles using a Thomas-Fermi potential and the WKB method. Moliere theory was further improved by Bethe, who included treatment for the scattering off of atomic electrons~\cite{Bethe:1953va}. Although Moliere theory provided a more rigorous treatment, Eq.~\ref{Rossi_MS} remained as the widely used approach due to the utility of a quickly-computable Gaussian approximation.

The prominence of Eq.~\ref{Rossi_MS} lead Highland to modify it for empirically better agreement with Moliere Theory~\cite{HIGHLAND1975497}. Highland replaced the \SI{14.8}{MeV} constant with the parameterization $S_2(1+\varepsilon \ln{x/X_o})$, to obtain
\begin{equation}
 \sigma_\psi^{ \rm plane} = \frac{1}{\sqrt{3}} \frac{S_2} { \beta c p} \sqrt{\frac{x}{X_o}} \left[ 1 + \varepsilon \ln{\frac{x}{X_o}}  \right].
 \label{Highland}
\end{equation}
Highland suggested setting $S_2 =$ \SI{13.9}{MeV} and $\varepsilon = 0.048$ to obtain the best fit to (Gaussian Approximations of) Moliere theory for $\beta=1$ particles~\cite{HIGHLAND1975497}.

Lynch and Dahl later noted that Highland did not use Bethe's prescription of Moliere Theory~\cite{Bethe:1953va}. They re-fit highland's parameters to angular distributions obtained from Bethe's prescription of Moliere scattering for heavy particles, as simulated in GEANT3~\cite{Brun:1987ma}. Since Moliere theory provides non-Gaussian angular distributions, Lynch and Dahl truncated their distributions, fitting the central 98\% to a Gaussian. Lynch and Dahl proposed the new fit values of $S_2 =$ \SI{13.6}{MeV} and $\varepsilon = 0.038$~\cite{LYNCH19916}. Furthermore, they included treatment for multiply charged particles and corrected the $\beta$ dependence to obtain 
\begin{equation}
 \sigma_\psi^{ \rm plane} = \frac{z}{\sqrt{3}} \frac{ \SI{13.6}{MeV} } { \beta c p} \sqrt{\frac{x}{X_o}} \left[ 1 + 0.038 ln(\frac{x z^2}{X_o \beta^2})  \right].
 \label{PDG_MS}
\end{equation}
Above, $z$ is the charge of the particle. Currently, the Particle Data Group (PDG) review on the passage of particles through matter quotes Eq.~\ref{PDG_MS} in the section on Multiple scattering through small angles~\cite{PhysRevD.110.030001}. It should be noted that although Ref.~\cite{PhysRevD.110.030001} presents Eq.~\ref{PDG_MS} as a general multiple Coulomb scattering formula, it is a fit to Moliere theory for heavy particles.

In our presentation of Eqs.~\ref{Rossi_MS}--\ref{PDG_MS}, we make use of the conversions~\cite{PhysRevD.110.030001}
\begin{equation}
    \sigma_\psi^{ \rm plane} = \frac{1}{\sqrt{3}} \sigma_\theta^{ \rm plane}
\end{equation}
and 
\begin{equation*}
    \sigma_\psi^{ \rm 3D} = \sqrt{2} \sigma_\psi^{ \rm plane},
\end{equation*}
which hold for small angles. Above, $\psi$ and $\theta$ are the angles illustrated in Fig.~\ref{fig:MS_pic}.

In Eqs.~\ref{Rossi_MS}--\ref{PDG_MS}, the variable $x$ represents the thickness of the scattering medium ($t$ in Fig.~\ref{fig:MS_pic}). This is because Refs.~\cite{RevModPhys.13.240,Bethe:1953va,HIGHLAND1975497,LYNCH19916} are concerned with scattering through a film of known thickness. However, earlier editions of the PDG~\cite{RevModPhys.45.S1} parameterized the angular resolution due to multiple scattering in terms of the length inside the scattering medium ($L$ in Fig.~\ref{fig:MS_pic}). In the most widely accepted prescription of Moliere theory~\cite{Bethe:1953va}, $t$ is commonly understood to represent the thickness of the scattering medium, but it is actually defined as the distance traveled by the scattering particle, which corresponds more closely to $L$ than $t$ in Fig.~\ref{fig:MS_pic}. The author refers to the difference between the thickness of the scattering medium and the distance traveled by the scattering particle as the ``detour factor'' and makes no attempt to take it into account.

\begin{figure}[htb]
\subfloat[\label{fig:degrad:a}]{%
  \includegraphics[width=0.45\textwidth,trim={3cm 0cm 1cm 1cm},clip]{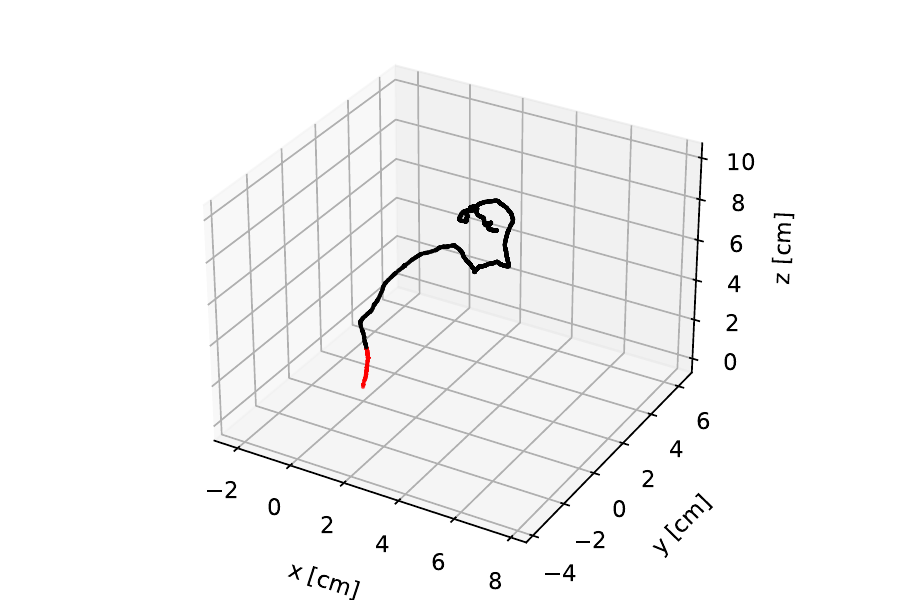}%
}\hfill\\
\subfloat[\label{fig:degrad:d}]{%
  \includegraphics[width=0.45\textwidth,trim={3cm 0cm 1cm 1cm},clip]{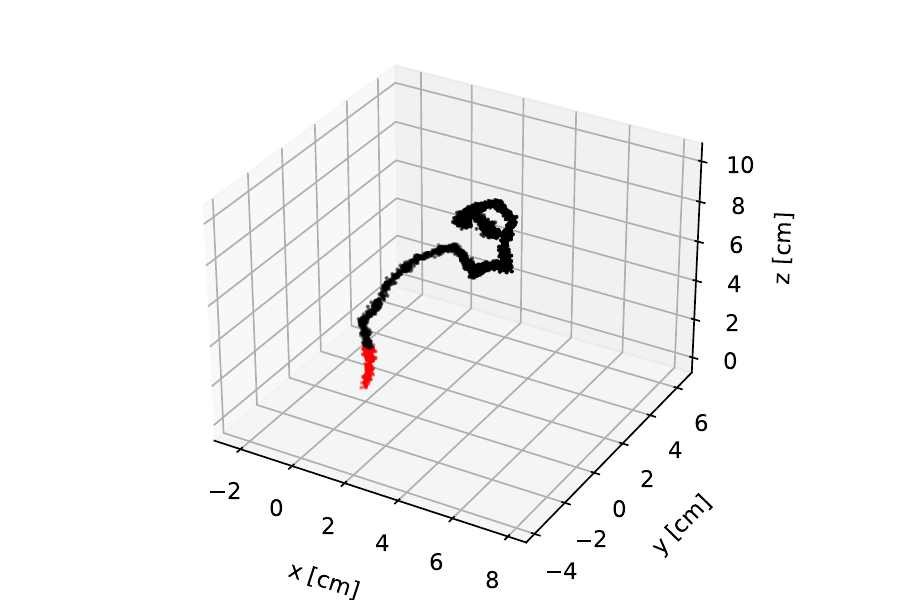}%
}\hfill
\caption{\label{fig:degrad} \texttt{Degrad} simulations of a $150$ keV electron in 60\% He : 40\% $\textrm{CF}_4$ at $20$ Degrees Celsius and $760$ torr. Subplot (a) is a raw \texttt{Degrad} simulation and subplot (b) is the same simulation after Gaussian smearing is applied, corresponding to 25 cm of drift in the gas mixture. A fit length of $x =$ \SI{2}{cm} is shown in red for both cases.}
\end{figure}

Since the detectors considered in this study can measure $L$, we adopt $L$ rather than $t$ as $x$. In the following discussion, $x$ will refer to the point-to-point length of the portion of the electron’s track used to determine its initial direction, and we refer to this as the fit length. As such, $x$ is a free parameter that can range from $0$ to the full point-to-point length of the recoil track. In practice, $x$ is typically much smaller than the total track length, as directionality is lost when the electron loses energy. An example is shown in Fig.~\ref{fig:degrad} where a fit length of $x =$ \SI{2}{cm} is indicated in red.

\subsection{Simulations}
\label{sims1}
While Moliere Theory and Eq.~\ref{Rossi_MS} are generally applicable, Eq.~\ref{PDG_MS} is a fit to Moliere theory for heavy particles, meaning it may not be accurate for electrons. Here, we follow a process similar to that of Ref.~\cite{LYNCH19916}, but our goal is to obtain a simple Gaussian approximation for the angular resolution of electrons in gas. For this purpose, we use \texttt{Degrad}~\cite{degrad}, which is recognized as the most complete simulation software for electron scattering in gas~\cite{degradbest}.

Using \texttt{Degrad}, we simulate 10,000 electron tracks for each energy and gas listed in Table~\ref{table1}. For each electron track, \texttt{Degrad} provides the $\{x_i, y_i, z_i, t_i\}$ coordinates for the ionized electrons, as illustrated in Fig.~\ref{fig:degrad:a}. All tracks begin at the origin with an initial direction of $\hat{z}$. We want to obtain $\psi$, as illustrated in Fig.~\ref{fig:MS_pic}. To do so, we loop through the time-ordered ionized electron coordinates and identify the first electron whose distance from the origin is greater than the predefined fit length ($x$). The fit length values explored are summarized in Table~\ref{table1}. The vector from the origin to this identified electron is projected onto the $yz$-plane and $\psi$ is the angle between the projected vector and $\hat{z}$. This process is repeated over all simulated tracks to generate a distribution of $\psi$ for each gas, energy, and fit length combination. Following Ref.~\cite{LYNCH19916},  we trim the angular distribution, keeping only the central 98\%, and fit that to a Gaussian to obtain the simulated value for $ \sigma_\psi^{ \rm plane}$ for each case. Since this treatment is for small angles, all cases satisfy $\sigma_\psi^{\rm plane} < 15^{\circ}$.

\begin{table}[b]
\caption{\label{table1}%
The cases for which we find the simulated values of $ \sigma_\psi^{\rm plane}$. All gasses are at \SI{760}{torr} and \SI{20}{\degreeCelsius}. The radiation length, $X_o$, is obtained from Ref.~\cite{AtomicNuclearProperties}. For each fit length, $x$, the energy range is selected so that $\sigma_\psi^{\textrm{plane}} < \ang{15}$. The energies are explored in steps of \SI{10}{keV} for energies less than 100 keV and steps of \SI{100}{keV} otherwise.}
\begin{ruledtabular}
\begin{tabular}{cccc}
\multicolumn{1}{c}{\textrm{Gas}}&{$\textrm{X}_o$ [m]}&{\textrm{x[cm]}}& {\textrm{ Energy [keV]}}\\
\colrule
 & & 0.5 & 200 -- 1000\\
$\textrm{CF}_4$ & 89.9 & 1.0 & 200 -- 1000\\
 & & 2.0 & 300 -- 1000\\
\hline
 & & 0.5 & 100 -- 1000\\
$\textrm{CO}_2$ & 197 & 1.0 & 200 -- 1000\\
 & & 2.0 & 200 -- 1000\\
\hline
 & & 0.5 & 60 -- 1000\\
$\textrm{CH}_4$ & 697 & 1.0 & 80 -- 1000\\
 & & 2.0 & 200 -- 1000\\
\hline
 & & 0.5 & 70 -- 1000\\
$\textrm{C}_2\textrm{H}_6$ & 362 & 1.0 & 100 -- 1000\\
 & & 2.0 & 200 -- 1000\\
\hline
 & & 0.5 & 60 -- 1000\\
$\textrm{Ne}$ & 345 & 1.0 & 90 -- 1000\\
 & & 2.0 & 200 -- 1000\\
\hline
 & & 0.5 & 100 -- 1000\\
$\textrm{Ar}$ & 118 & 1.0 & 200 -- 1000\\
 & & 2.0 & 300 -- 1000\\
\hline
 & & 0.5 & 300 -- 1000\\
$\textrm{Xe}$ & 15.5 & 1.0 & 500 -- 1000\\
 & & 2.0 & 800 -- 1000\\
\end{tabular}
\end{ruledtabular}
\end{table}

\subsection{Results}
\label{MSresults}
Similarly to Ref.~\cite{LYNCH19916}, we fit the $S_2$ and $\varepsilon$ parameters of Eq.~\ref{Highland} to the simulated data sets summarized in Table~\ref{table1}. A simultaneous fit to all cases yields  $S_2 =$ \SI{7.57}{MeV} and $\varepsilon = -0.078$, which we refer to as Fit I. One issue with the parameterization of Eq.~\ref{Highland} is that when $\varepsilon \neq 0$, it is possible to create physically viable scenarios where the equation predicts $\sigma_\psi^{ \rm plane} < 0$. To address this, we perform another fit using the original form of Eq.~\ref{Rossi_MS} by setting $\varepsilon = 0$. This second fit, referred to as Fit II, gives $S_2 =$ \SI{13.2}{MeV}.

\begin{figure}[b]
\includegraphics[width=0.45\textwidth]{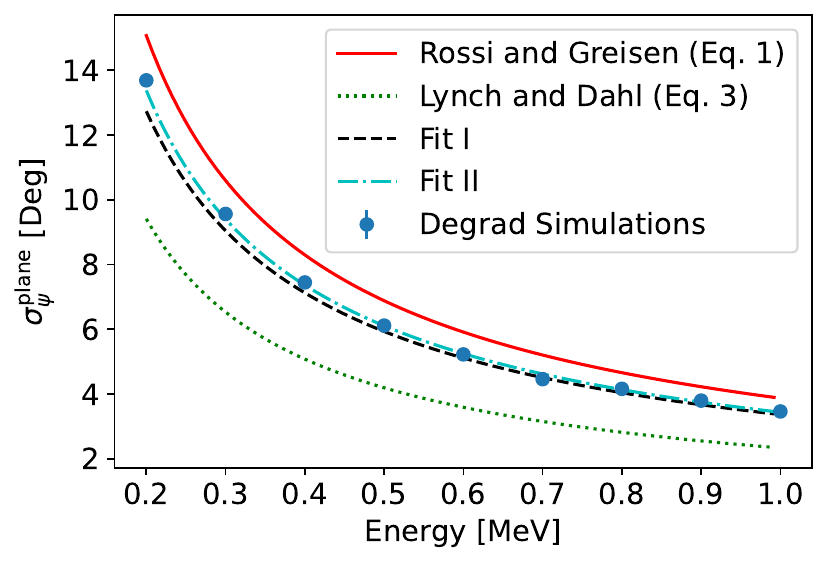}
\caption{\label{fit_fig} Angular resolution versus energy for electrons in $\textrm{CF}_4$ at \SI{760}{torr} and \SI{20}{\degreeCelsius}. A fit length of $x=$ \SI{1.0}{cm} is used throughout. The blue points are the results from \texttt{Degrad} simulations, described in Section~\ref{sims1}. Fits I and II are plotted alongside Eqs.~\ref{Rossi_MS} and~\ref{PDG_MS}. }
\end{figure}

An example of our fits, shown alongside the simulation results for $\textrm{CF}_4$ for a fit length of \SI{1.0}{cm}, is presented in Fig.~\ref{fit_fig}. For comparison, we have also plotted Eqs.~\ref{Rossi_MS} and~\ref{PDG_MS} in the same figure. Since Eq.~\ref{Rossi_MS} is a general approximation, while Eq.~\ref{PDG_MS} is based on a fit for heavy particles, it is not surprising that the former provides better predictions with respect to our simulations. Both Fit I and Fit II represent the simulation data more closely then the original formulae.

To compare Fit I and Fit II, we evaluate them on an independent test set. This test set simulates two new gas mixtures: 70\% He : 30\% $\textrm{CO}_2$ and 60\% He : 40\% $\textrm{CF}_4$, both at \SI{760}{torr} and \SI{20}{\degreeCelsius}. Testing on gas mixtures that were not used to obtain Fits I and II allows us to evaluate their generalizability. The details of the test simulation set are provided in Table~\ref{table2}.

\begin{table}[b]
\caption{\label{table2}%
The cases simulated to test Fit I and Fit II. All mixtures are at $760$ torr and \SI{20}{\degreeCelsius}. The radiation length, $X_o$, is computed for each mixture as described in Ref.~\cite{PhysRevD.110.030001}. The energy range is selected so that $\sigma_\psi^{\textrm{plane}} < \ang{15}$. The energies are explored in steps of \SI{10}{keV}.
}
\begin{ruledtabular}
\begin{tabular}{cccc}
\multicolumn{1}{c}{\textrm{Gas}}&{$\textrm{X}_o$ [m]}&{\textrm{x [cm]}}& {\textrm{Energy Range [keV]}}\\
\colrule
70\% He : 30\% $\textrm{CO}_2$ & 606 & 0.5 & 50 -- 200\\
60\% He : 40\% $\textrm{CF}_4$ & 220 & 0.5 & 70 -- 200\\
\end{tabular}
\end{ruledtabular}
\end{table}

\begin{figure}[b]
\includegraphics[width=0.45\textwidth]{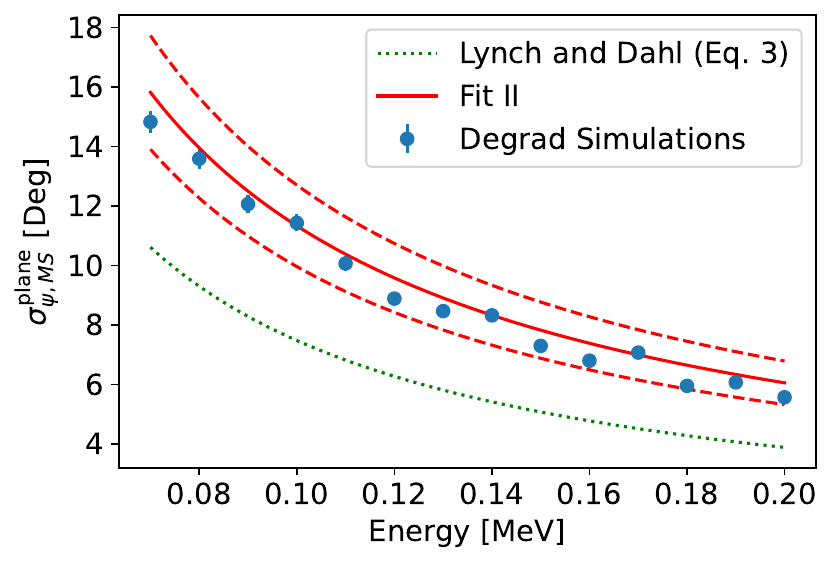}
\caption{\label{test_fig} 
Angular resolution versus energy for electron recoils in 60\% He : 40\% $\textrm{CF}_4$ at \SI{760}{torr} and \SI{20}{\degreeCelsius}. A fit length of $x=$ \SI{0.5}{cm} is used throughout. The blue points are the results from \texttt{Degrad} simulations, described in Section~\ref{sims1}. The red line is Eq.~\ref{final_MS}, the dashed red lines indicate the uncertainty on Eq.~\ref{final_MS} and the dotted green line is Eq.~\ref{PDG_MS}.}
\end{figure}

The angular resolution obtained from simulations for the test cases in Table~\ref{table2} is compared to the predictions from Fit I and Fit II. The Root Mean Square Error (RMSE) between the predictions and the simulation results are \ang{0.534} and \ang{0.396}, respectively. Fig.~\ref{test_fig} shows an example of Fit II plotted alongside the \texttt{Degrad} simulation results for 60\% He : 40\% $\textrm{CF}_4$. Given that Fit II demonstrates better performance on the test set, we adopt it as our Gaussian approximation for the angular resolution due to multiple scattering of electrons in gas. Our final multiple scattering approximation is
\begin{equation}
 \sigma_{\psi, \textrm{MS} }^{ \rm plane} = \frac{1}{\sqrt{3}} \frac{13.2 \pm 1.59 {\, \rm MeV}} { \beta c p} \sqrt{\frac{x}{X_o}}.
 \label{final_MS}
\end{equation}
Above, 1.59 MeV is the systematic uncertainty on $S_2$, estimated by fitting $S_2$ for each gas in Table~\ref{table1}, individually, then taking the standard deviation of all fit values.

As it stands, Eq.~\ref{final_MS} provides an estimate of the multiple scattering contribution to the angular resolution as a function of fit length ($x$). Consequently, Eq.~\ref{final_MS} suggests that an arbitrarily small angular resolution can be achieved by using a small enough fit length. Of course this result is not particularly useful for estimating the best achievable angular resolution of a detector, to do so we must include the effective point resolution of the detector.

\section{The Effective Point Resolution}
\label{PR}

\subsection{Background}
\label{diffBG}

The relationship between the effective point resolution and the angular resolution has already been studied in the context of gaseous TPCs. In Ref.~\cite{Vahsen:2014fba}, it is found that if $N$ points are placed uniformly in a line of length $x$ and measured with an effective resolution of $\sigma_{x,y,z}$, the angular resolution is 
\begin{equation}
    \label{pointres}
    \sigma_{\psi, \textrm{PR} }^{\textrm{plane}} = \frac{\sqrt{12} \sigma_{x,y,z}}{x\sqrt{N}}.
\end{equation}
In TPCs, $\sigma_{x,y,z}$ contains contributions both from the diffusion of the secondary electrons as they drift in the TPC, and from the position resolution of the TPC readout plane. Treating the multiple scattering and effective point resolution as two independent effects, we approximate the overall angular resolution as the sum in quadrature
\begin{align}
    \sigma_{\psi, \textrm{total} }^{ \rm plane} &= \sqrt{(\sigma_{\psi, \textrm{MS} }^{\textrm{plane}} )^2 + (\sigma_{\psi, \textrm{PR} }^{\textrm{plane}} )^2} \nonumber  \\
    &= \sqrt{ \frac{1}{3} \frac{(13.2 \pm 1.59 {\, \rm MeV})^2} { \beta^2 c^2 p^2} \frac{x}{X_o}  + \frac{12 \sigma_{x,y,z}^2}{x^2 N} }.
    \label{RMS_tot1}
\end{align}
Above we take $N$ to be the number of ionized electrons within the fit length, $x$. This amounts to assuming a detector capable of counting individual electrons. Current MPGD-based TPCs are approaching this limit~\cite{Ligtenberg:2020ofy}, hence Eq.~\ref{RMS_tot1} estimates the best achievable angular resolution for a given fit length.

Equation~\ref{pointres} assumes that the ionized electrons are uniformly spread throughout the fit length, which means that the fractional energy loss of the primary electron is small and the diffusion of the secondary electrons is large compared to the spacing of ionization clusters. Following this assumption, the number of ionized electrons may be approximated as
\begin{equation}
    N=\frac{x dE/dx}{W},
    \label{N}
\end{equation}
where $W$ is the average energy to create and electron-ion pair and $dE/dx$ is the energy-loss per unit length. We re-package Eq.~\ref{RMS_tot1} by defining 
\begin{equation*}
    a \equiv \frac{1}{\sqrt{3}} \frac{13.2 \pm 1.59 {\rm \, MeV}}{\beta c p \sqrt{X_o}},
    \label{a}
\end{equation*}
and
\begin{equation*}
    b \equiv \sigma_{x,y,z}  \sqrt{\frac{12W}{dE/dx}} .
    \label{b}
\end{equation*}
to obtain the simple expression
\begin{equation}
    \sigma_{\psi, \textrm{total} }^{ \rm plane}  = \sqrt{a^2 x + b^2 x^{-3}} .
    \label{Rms_final}
\end{equation}
The constants $a$ and $b$ can be calculated from the electron recoil energy and basic detector properties ($X_o$, $W$, $dE/dx$, $\sigma_{x,y,z}$).

Since Eq.~\ref{Rms_final} accounts for both multiple scattering and effective point resolution, it exhibits a global minimum, representing the optimal balance between these two effects. This removes the ambiguity regarding the choice of fit length, which should be selected to minimize Eq.\ref{Rms_final}. The predicted optimal fit length is
\begin{equation}
    x_o = \frac{3^{ \frac{1}{4} }b^{ \frac{1}{2} }}{a^{ \frac{1}{2} }}.
    \label{opt_l}
\end{equation}
Substituting $x_o$ into Eq.~\ref{Rms_final}, we obtain:
\begin{align}
    \sigma_{\psi, \textrm{total} }^{ \rm plane} (x_o) &= \frac{2 a^{\frac{3}{4}} b^{ \frac{1}{4}} }{ 3^{ \frac{3}{8} } }\\ 
    &= \frac{2^\frac{5}{4} (13.2 \pm 1.59 {\, \rm MeV})^\frac{3}{4} \sigma_{x,y,z}^\frac{1}{4} W^\frac{1}{8} }{3^\frac{5}{8} (\beta cp)^\frac{3}{4} X_o^\frac{3}{8} \frac{dE}{dx}^\frac{1}{8}}
    \label{best_ang_res}
\end{align}
This is an approximation of the best achievable angular resolution given the electron recoil energy and detector properties ($X_o$, $W$, $dE/dx$, $\sigma_{x,y,z}$). One could use Eq.~\ref{best_ang_res} to optimize gas TPCs specifications for the angular resolution of electron recoils at an energy of interest; or conversely, to approximate angular resolution versus energy given the detector specifications.

\subsection{Simulations}
\label{diff_sims}
To evaluate the framework developed in Section~\ref{diffBG}, we incorporate an effective point resolution into the test set, summarized in Table~\ref{table2}. For the 60\% He : 40\% $\textrm{CF}_4$ gas mixture, we assume a \SI{600}{V/cm} drift field, consistent with Ref.~\cite{Baracchini:2020btb}. The corresponding diffusion coefficients are obtained as $\sigma_{ \rm T} =$ \SI{136}{\micro m / \sqrt{cm}} and $\sigma_{ \rm L} =$ \SI{114}{\micro m / \sqrt{cm}} using \texttt{Magboltz}~\cite{magboltz}. Since our framework considers only isotropic effective point resolution, we use the larger value of \SI{136}{\micro m / \sqrt{cm}}. Assuming a drift length of \SI{25}{cm}, the diffusion is estimated as $136 \cdot \sqrt{5}=$\SI{680}{\micro m}. Further assuming a \SI{100}{\micro m} readout segmentation, the binning contribution to the effective point resolution is $100/\sqrt{12} =$\SI{28.9}{\micro m}. The effective point resolution is then estimated as $\sigma_{x,y,z} =$ \SI{681}{\micro m}, the quadrature of the two effects. Therefore, \SI{681}{\micro m} of Gaussian smearing is applied to the all simulations in the 60\% He : 40\% $\textrm{CF}_4$ test case. For the 70\% He : 30\% $\textrm{CO}_2$ test case, \SI{466}{\micro m} of Gaussian smearing is applied, which is the largest expected point resolution from a detailed study of the BEAST TPCs~\cite{Jaegle:2019jpx}.
An example of a simulated electron track before and after Gaussian smearing of the secondary electrons is shown in Fig.~\ref{fig:degrad}.

To determine $\psi$ for each simulation after Gaussian smearing, we first isolate all points within a fit length $x$ from the track's starting point. Singular value decomposition (SVD) is used to identify the principal axis of this set of points, and a direction is assigned to ensure a positive dot product with $\hat{z}$ (the initial direction). This procedure is consistent with the Best-Expected method in Ref.~\cite{Ghrear:2024rku}. The remaining steps to calculate $\sigma_\psi^{\rm plane}$ follow the procedure detailed in Section~\ref{sims1}.

\subsection{Results}

\begin{figure}[b]
\includegraphics[width=0.45\textwidth]{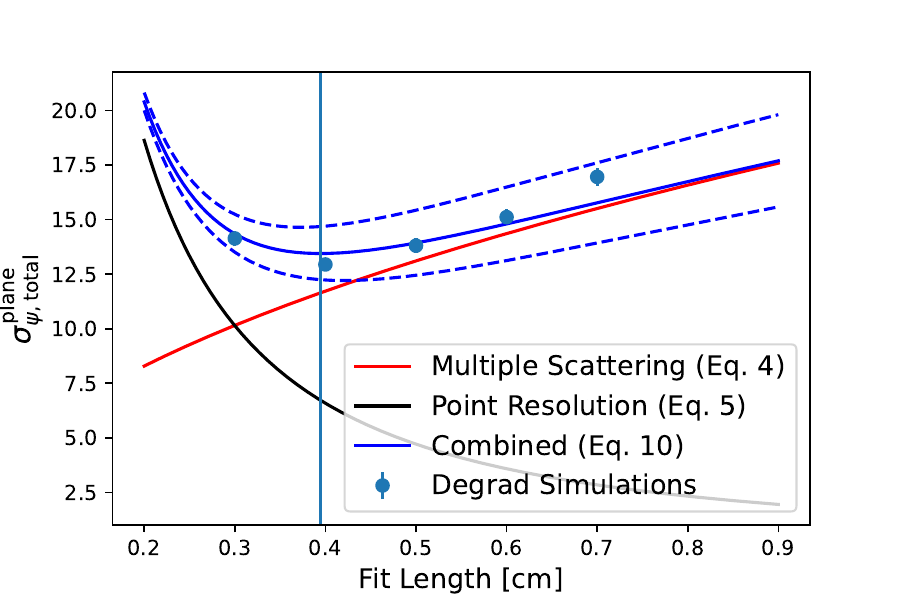}
\caption{\label{final_fig} Angular resolution versus fit length for \SI{50}{keV} electron recoils in 70\% He : 30\% $\textrm{CO}_2$ at \SI{760}{torr} and \SI{20}{\degreeCelsius} with \SI{466}{\micro m} Gaussian smearing. The blue points are the simulation results. The predicted multiple scattering contribution (red), point resolutions contribution (black), and the combined angular resolution (blue) are also plotted. The vertical line is the optimal length, $x_o$.}
\end{figure}

To apply the framework developed in Section~\ref{diffBG}, we first determine the parameters $a$ and $b$. First, $a$ can be readily computed using the electron recoil energy and $X_o$, both provided in Table~\ref{table2}. Calculating $b$ requires knowledge of $\sigma_{x,y,z}$, $W$, and $dE/dx$. The values of $\sigma_{x,y,z}$ are specified in Section~\ref{diff_sims}. There is extensive literature on $dE/dx$ and $W$. Depending on the application, it may be suitable to obtain their values via theoretical calculation~\cite{bloch1933bremsung,PhysRevD.110.030001}, previous experimental measurements~\cite{berger1982stopping,6bb86445}, or simulation tools~\cite{degrad,veenhof1998garfield}. Here, we obtain self-consistent estimates of $dE/dx$ and $W$ by using \texttt{Degrad}. The goal is to demonstrate that our framework is effective when $dE/dx$ and $W$ can be accurately estimated.

Since only the fraction $\frac{W}{dE/dx} \equiv <dN/dx>$ appears in $b$, we calculate $<dN/dx>$ directly. Assuming negligible energy loss across the fit length we compute a single value of $<dN/dx>$ for each electron energy / gas mixture. For each simulated electron track at a given energy and gas mixture, we plot the number of ionized electrons versus the fit length up to \SI{0.4}{cm}. This plot is fitted with a line passing through the origin, yielding a slope $dN/dx$ and $<dN/dx>$ is obtained as the average over all simulated tracks.

 In Fig.~\ref{final_fig}, we compare the predicted angular resolution from Eq.~\ref{Rms_final} to the results from simulation described in Section~\ref{diff_sims} for $50$\,keV electrons in 70\% He : 30\% $\textrm{CO}_2$. The predicted and observed angular resolutions agree well, particularly at their minima. At longer fit lengths and larger angles, our approximations begin to break down, resulting in reduced predictive accuracy. Note that the predicted optimal fit length ($x_o$), shown as a vertical line in the figure, effectively coincides with the minimum in the observed angular resolution from the \texttt{Degrad} simulations. 

\begin{figure}[b]
\includegraphics[width=0.45\textwidth]{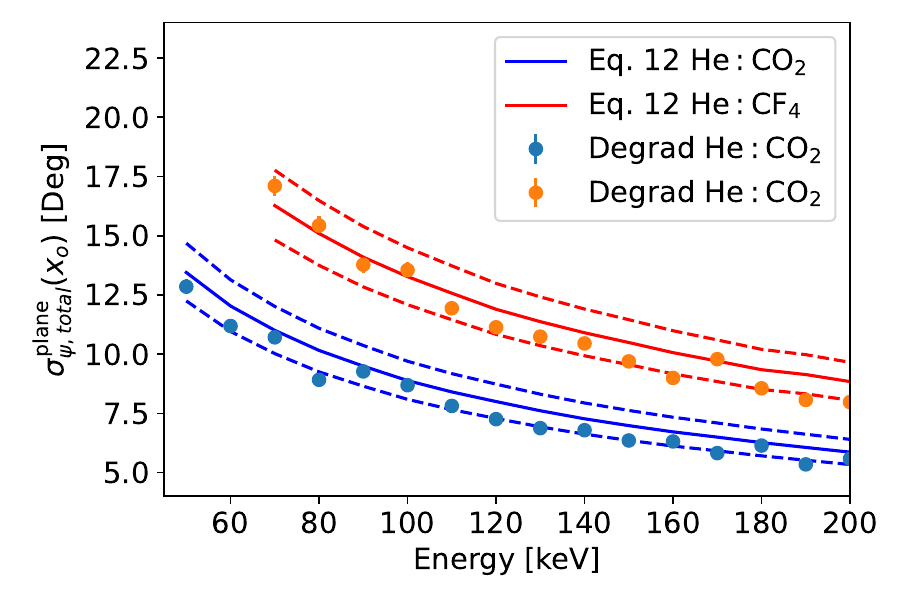}
\caption{\label{final_fig2} Angular resolution versus energy using a fit length of $x=x_o$ (the optimal fit length as calculated in Eq.~\ref{opt_l}). The points are the simulation results and the solid lines Eq.~\ref{best_ang_res} for  70\% He : 30\% $\textrm{CO}_2$ (orange) and 60\% He : 40\% $\textrm{CF}_4$ (blue). Both gas mixtures are at \SI{760}{torr} and \SI{20}{\degreeCelsius}.}
\end{figure}

Finally, we test Eq.~\ref{best_ang_res} by comparing it's predictions for the best achievable angular resolution to the observed angular resolution in simulation when a fit length of $x=x_o$ is used. This is done for all test cases in Table~\ref{table2} and the results are presented in Fig.~\ref{final_fig2}. The mean error is \ang{0.59}.

\section{Conclusion}

In this work, we investigated the angular resolution of electron recoils in gaseous detectors. First, we demonstrated that the widely used formula for multiple scattering in the PDG is a fit to Moliere theory for heavy particles and, as such, is not directly applicable to electrons. To rectify this, we obtained a modified expression that accurately describes multiple scattering for electrons in gaseous media.

In addition, we incorporated the effects of secondary-electron diffusion and detector resolution to develop a more complete model for estimating the best achievable angular resolution of electrons in gas. This model identifies an optimal fit length--balancing multiple scattering and spatial resolution effects--and provides a simple, practical formula to estimate the corresponding optimal angular resolution. The best  achievable angular resolution can be estimated using only information about the electron energy, and basic gas and detector properties ($W$, $dE/dX$, $X_o$, $\sigma_{x,y,z}$).

The approach was validated through detailed \texttt{Degrad} simulations, which showed good agreement with our predictions across different gas mixtures and experimental conditions. The results provide a useful tool for optimizing gaseous detectors for electron directionality, offering insights into gas selection, detector design, and data analysis strategies. Given the broad applications of gaseous detectors in many areas of physics, this work is relevant to multiple scientific communities.

\begin{acknowledgments}
This work was supported by the U.S. Department of Energy (DOE) via Award Number DE-SC0010504.
\end{acknowledgments}

\appendix

\bibliography{apssamp}

\end{document}